\documentclass[aps,onecolumn,amssymb]{revtex4}
\usepackage{graphicx}

\def\be{\begin{equation}}
\def\ee{\end{equation}}
\begin{document}

 \title{Critical entanglement of XXZ Heisenberg chains with defects}
 \author{Jize Zhao$^1$, Ingo Peschel$^{2,3}$ and Xiaoqun Wang$^{1,4}$\\
 {\small $^1$Institute of Theoretical Physics, Chinese Academy of Sciences, 
 Beijing 100080, China}\\
 {\small $^2$Max-Planck-Institut f\"ur Physik komplexer Systeme,
 N\"othnitzer Strasse 38, D-01187 Dresden, Germany}\\
 {\small $^3$ Fachbereich Physik, Freie Universit\"at Berlin, Arnimallee 14, D-14195 Berlin,
 Germany}\\
 {\small $^3$Department of Physics, Renmin University of China,
 Beijing 100872, China}}
 \begin{abstract}
 We study the entanglement properties of anisotropic open spin one-half Heisenberg chains
 with a modified central bond. The entanglement entropy between the
 two half-chains is calculated with the density-matrix renormalization method (DMRG).
 We find a logarithmic behaviour with an effective central charge
 $c'$ varying with the length of the system. It flows to one in the
 ferromagnetic region and to zero in the antiferromagnetic region of the model. In the
 XX case it has a non-universal limit and we recover previous results.

\end{abstract}
\maketitle

\vspace{1cm}

The consideration of entanglement properties has brought a new element
into the study of many-particle quantum states. Entanglement is related
to a division of the system into two parts and can be quantified via
the reduced density matrix $\rho$ and the 
entropy $S= -{\rm tr}(\rho \ln\rho)$ connected with it. This quantity is 
particularly interesting for the ground state of critical one-dimensional 
systems. For example,
if one cuts an open chain into two halves of length $L$, conformal invariance
predicts the universal form \cite{Holzhey94,Calabrese04}
\begin{equation}
S =\frac {c}{6}\ln L+k
\label{log}
\end{equation}
where $c$ is the central charge in the conformal classification. An analogous
formula with $c/6$ replaced by $c/3$ holds for a segment of length $L$
in an infinite chain. 
The logarithmic divergence is a particular signature of the criticality
and can be related to analogous universal contributions
to the free energy of critical two-dimensional systems with conical
shape \cite{Cardy-Peschel88,Calabrese04,Levine04}. It has been
verified numerically for a number of quantum chains
\cite{Vidal03,Latorre03} and
derived analytically for free fermions hopping on a chain \cite{Jin04}.
For this system it can also be proven by putting proper bounds on $S$
\cite{Wolf05,Gioev-Klich05}.

Since the entanglement entropy measures the mutual coupling of the
two parts of a system in the wave function, a defect at their boundary 
should have a strong influence. In one dimension, defects show particularly
interesting features in Luttinger liquids, i.e. in strongly correlated
fermionic systems. Their effective strength then depends on the sign of the 
interaction \cite{Kane-Fisher92,Eggert-Affleck92} and goes
to zero for attraction while it diverges for repulsion, as the system size
increases. This has been checked in various further studies, see e.g. 
\cite{Qin96,Qin97,Rommer00,Meden02,Louis03,Andergassen04}.
One then wonders how the entanglement behaves in such a case and 
how the logarithmic law (\ref{log}) is affected. This question was first 
raised by Levine \cite{Levine04} who used bosonization and found results to 
lowest order in the impurity strength which are consistent with the 
general picture. The simpler case of a free-fermion hopping model, corresponding
to the XX spin chain, was treated
numerically in Ref. \cite{Peschel05}. In this case, the logarithmic behaviour
was found to persist, but with a prefactor $c_{eff}$ which depends continuously
on the defect strength.

In the present paper we present a numerical study of this problem for the case 
of a planar XXZ spin chain, which has $c = 1$ and is the lattice version of a 
Luttinger model. We treat open chains with one modified bond in the middle and use
density-matrix renormalization (DMRG) \cite{White93,Book98}. This method
is ideally suited for such a study since the calculation of $\rho$ and
its spectrum is an intrinsic part of the algorithm. We find that the effective
central charge, which will be called $c'$, tends asymptotically to one for 
$\Delta < 0$ (ferromagnetic region, resp. attraction) and to zero for $\Delta > 0$
(antiferromagnetic region, resp. repulsion).
How fast this happens depends, however, on the defect strength, and for
weak perturbations and interactions the asymptotic region is still far
away, although we treated chains up to 800 sites. We also compared to Levine's
formulae but could not obtain a quantitative agreement, although we verify
his general picture.
In the following  we first present the model and the method of calculation.
Then we show some density-matrix spectra and their typical features.
After that we present the calculations of $S$ and the results for the
quantity $c'$ for a number of defect strengths $J_{imp}$ and
anisotropic parameters $\Delta$. Finally we compare with Ref.\cite{Levine04}
and draw our conclusions.\\

\section{Model and method}

We studied the XXZ model for an open chain of $2L-1$ sites, which is divided
into a left part with $L$ sites and a right part with $L-1$ sites. Both
are connected by a bond of strength $J_{imp}$ while for the rest of the
chain we take $J=1$. The Hamiltonian then has the form
\be
H=\sum_{i=1}^{L-1}h_{i}+J_{imp}\;h_{L}+\sum_{i=L+1}^{2L-2}h_{i}
\label{chain}
\ee
where
\be
h_{i}=\frac{1}{2}\left( {S}_{i}^{+}{S}_{i+1}^{-}+{S}_{i}^{-}{S}_{i+1}^{+}\right)
+\Delta S_{i}^{z}S_{i+1}^{z}
\label{H}
\ee
and the anisotropy $\Delta$ lies between $-1$ and $+1$. Written in terms of 
spinless fermions, the $S^zS^z$-term corresponds to a nearest-neighbour interaction
which is repulsive (attractive) for $\Delta > 0$ ($\Delta < 0$).
The system was studied in its ground state which in the present case corresponds 
to total spin $S^{z}=\pm 1/2$, and we took $S^{z}= +1/2$. In the fermionic
formulation, the filling is $L/(2L-1)$ and thus slightly above $1/2$.
We chose the odd total length in order to avoid relatively large oscillations 
between even and odd $L$-values which are present in even-length chains. 
These are connected with the different dimerization patterns 
in the spin correlation functions which are induced by the open ends.
Although this is an interesting topic, we preferred to avoid it for the present
purpose.\\

To determine the effect of the impurity bond on the entanglement entropy $S$ 
between left side and right side, we performed extensive DMRG calculations.
As usual in such calculations, we first determined the ground state wavefunction of
(\ref{chain}) and then traced out the spin degrees of freedom of the right side
to obtain the reduced density matrix $\rho$. The defect was inserted only at the
last step of the procedure. We worked with $m=500$ states and used no sweeps.
The values of $L$ varied between 3 and 401, corresponding to a total length of 
up to $801$ sites. The computations were performed on the HPSC45 of ICTS, CAS.\\

\begin{figure}
\includegraphics[width=6.8cm]{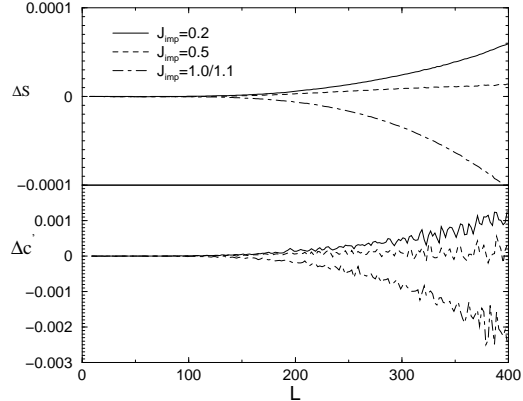}
\caption{Comparison between DMRG and correlation function results
for the XX case and $J_{imp}=0.2, 0.5$ and $1.0/1.1= 0.9091$. 
Upper panel: Difference in the entanglement entropy $S$.
Lower panel: Difference in the effective central charge $c'$. }
\label{BENCHMARK}
\end{figure}

To judge the accuracy of these calculations, we used the XX case, 
$\Delta=0$, where the system becomes a simple free-fermion hopping model. Then
one can determine $\rho$ from the one-particle correlation function $C_{ij}=
<c_i^{\dagger}c_j>$ by diagonalizing the matrix $C$ in the subsystem
\cite{Peschel03}. Such calculations were done recently for segments
in homogeneous chains with interface defects \cite{Peschel05} and in random chains
\cite{Laflorencie05}. For the present geometry with open ends, this correlation
function can be determined analytically for the pure system ($J_{imp}=1$). Then
one finds
\be
C_{ij}=\frac {1} {4L} \left[
\displaystyle{ \frac {\sin(\frac {\pi}{2} \frac {2L+1}{2L}(i-j))}
{\sin(\frac {\pi}{4L}(i-j))}}
 -\displaystyle{\frac {\sin(\frac {\pi}{2} \frac {2L+1}{2L}(i+j))}
{\sin(\frac {\pi}{4L}(i+j))}}\right]
\label{corr}
\ee
With a defect, $C_{ij}$ has to be calculated numerically using the single-particle
eigenfunctions of $H$ and the filling. The corresponding results were considered 
as benchmarks. In Fig.\ref{BENCHMARK} we show the difference between the
DMRG results and the correlation function results for three defect strengths.
Both $\Delta S$ and $\Delta c'$ increase with the chain length, as one would
expect. From the upper panel one sees that the error in the entropy becomes about 
$10^{-4}$ for $L=400$. The error in the effective central 
charge shown in the lower panel is larger and shows more noise. This is due to
the numerical derivative by which one calculates $c'$
\be
c'(L)= 6 \;\left [\frac {S(L+2)-S(L-2)}{\ln (L+2) -\ln(L-2)} \right ]
\label{ceff}
\ee
and in which small differences enter. Thus $\Delta c'$ reaches about $10^{-3}$ at
$L=400$, which is still sufficient for our purposes. We expect the same accuracy also 
for the interacting system with
$\Delta$ non-zero. Note that in (\ref{ceff}) steps of 2 are used, because there
still is a small even-odd oscillation in $S$. The two resulting values for $c'$ approach
each other for increasing $L$. In the following we always give
results obtained for odd $L$.\\ 

\section{Spectra}

\begin{figure}
\includegraphics[width=6.8cm]{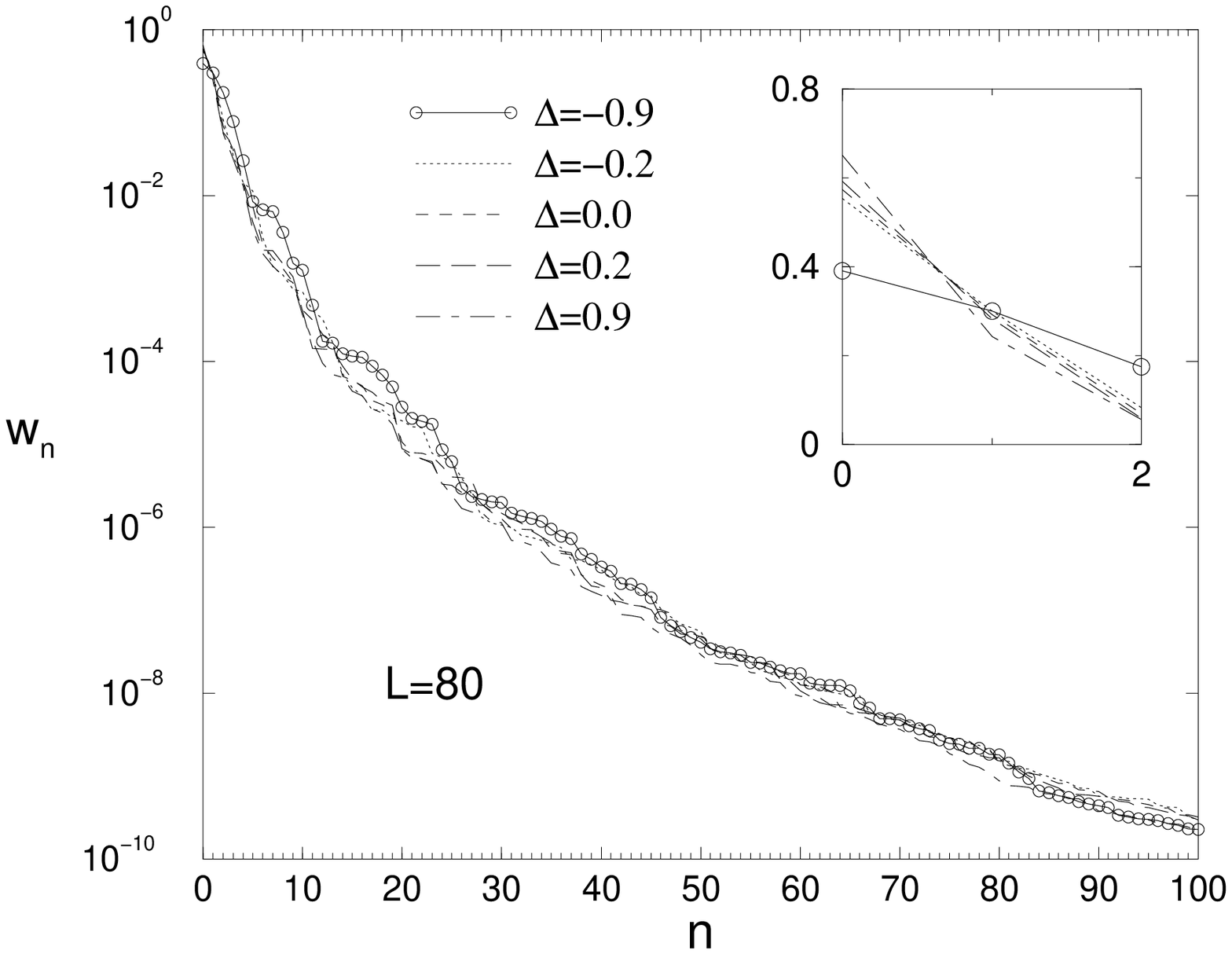}
\includegraphics[width=6.8cm]{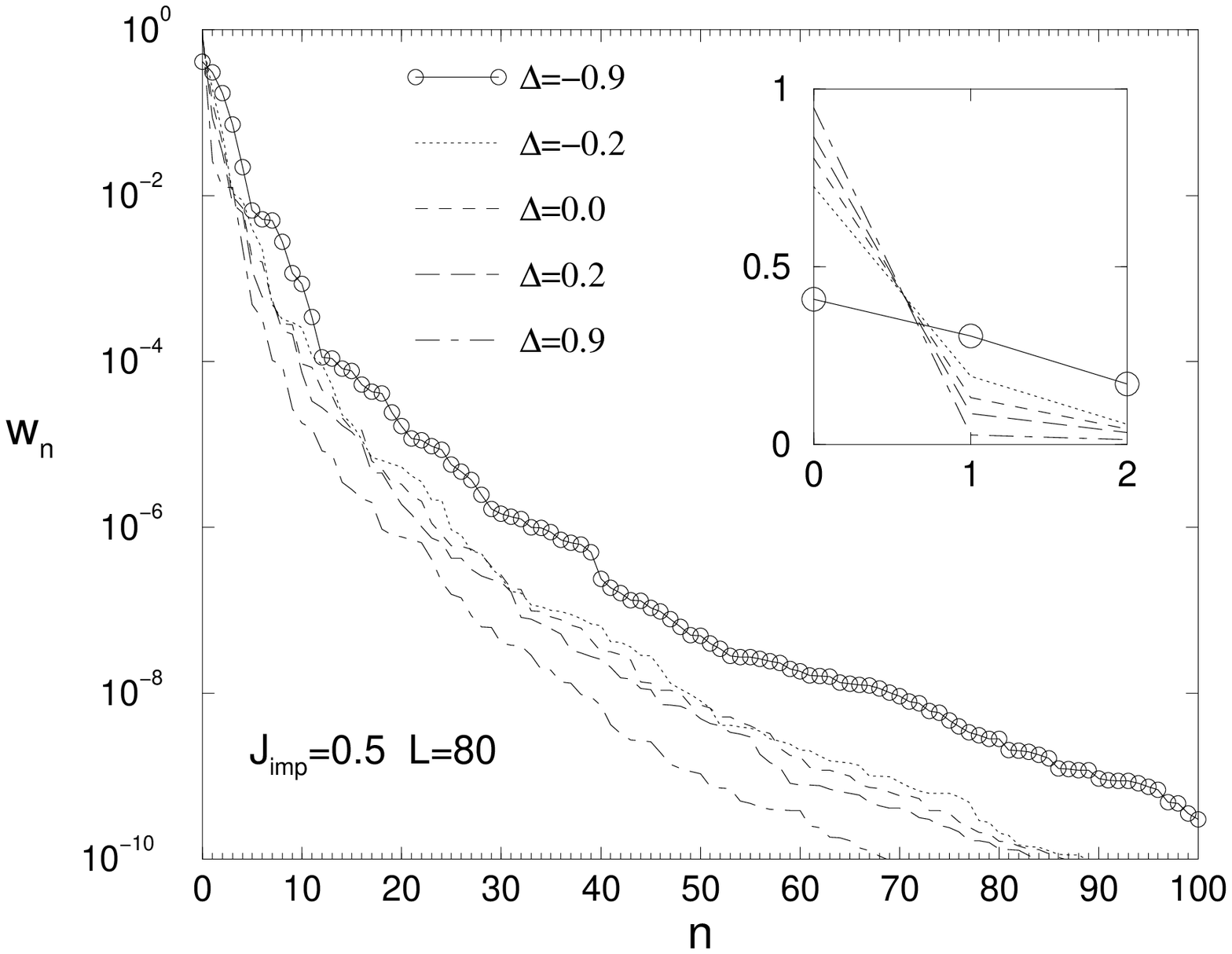}
\caption{Spectrum of the reduced density matrix for XXZ chains with various anisotropies.
Left: pure chains. Right: defect strength $J_{imp}= 0.5$.  Please note
the logarithmic scale. The insets show the first three eigenvalues.}
\label{SPEC1}
\end{figure}

\begin{figure}
\includegraphics[width=6.8cm]{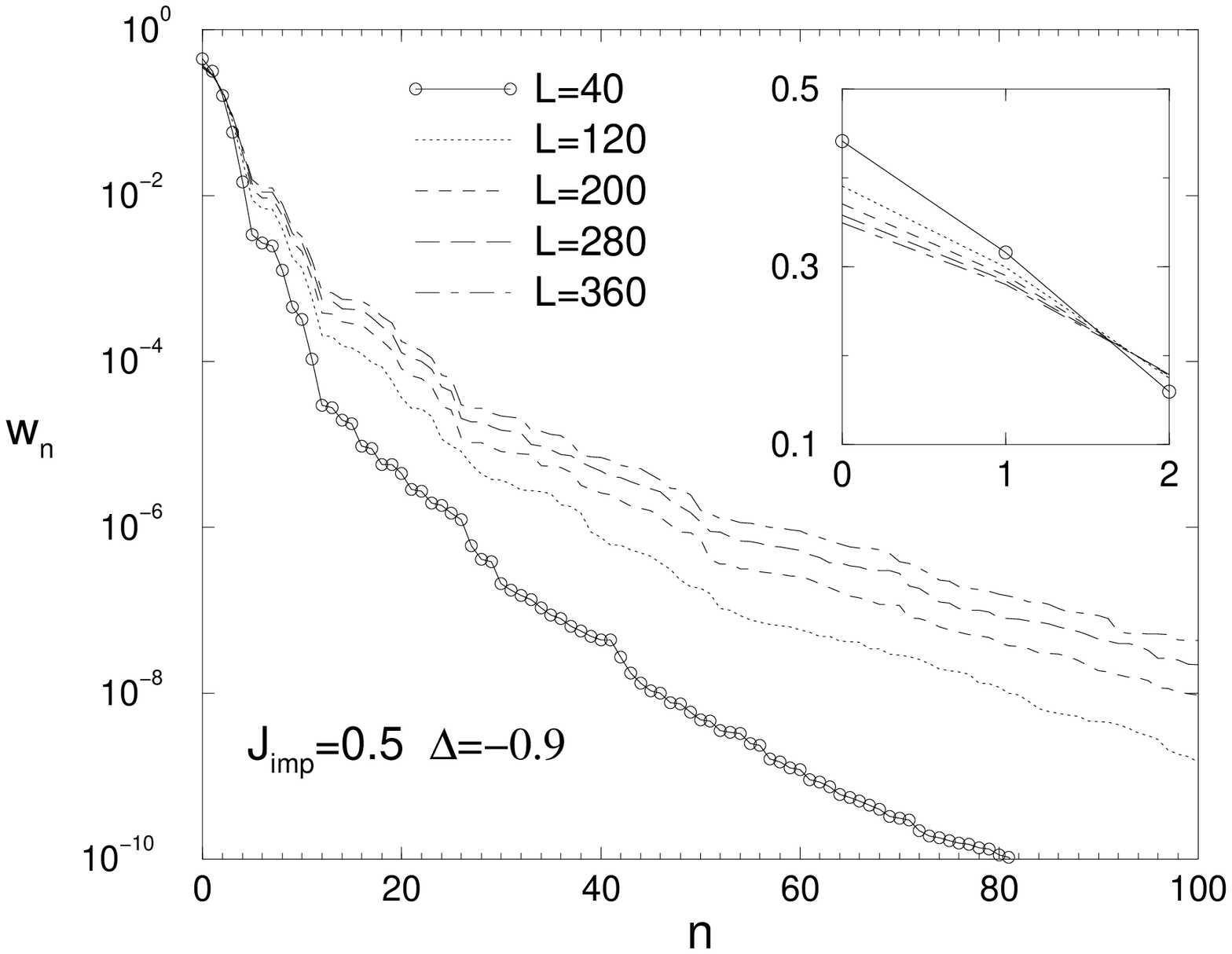}
\includegraphics[width=6.8cm]{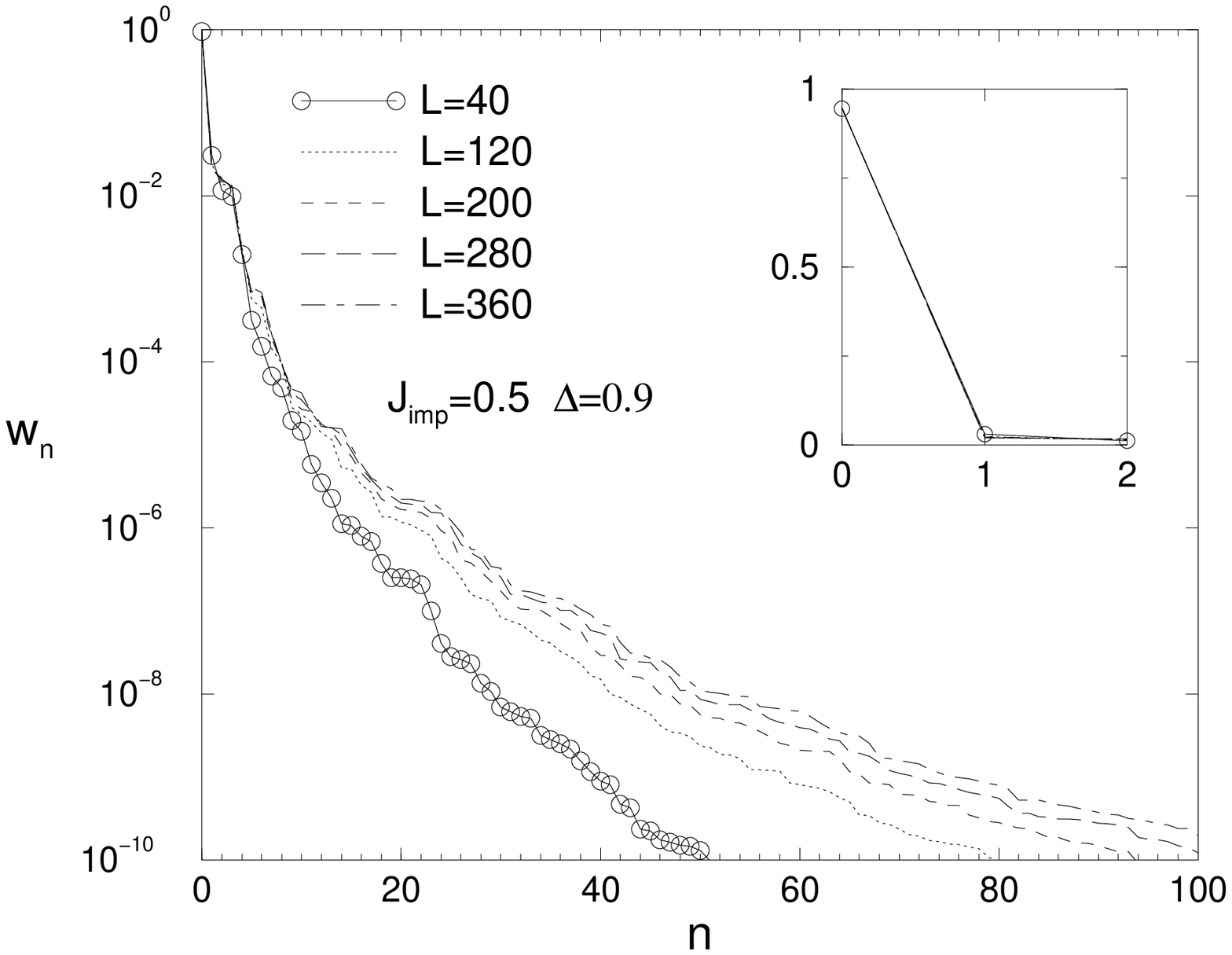}
\caption{Spectrum of the reduced density matrix for $J_{imp}=0.5$ 
and various sizes $L$.
Left: $\Delta= -0.9$. Right: $\Delta= +0.9$.
The insets show the first three eigenvalues.}
\label{SPEC2}
\end{figure}

Before we present the results for the entanglement entropy we show some 
density-matrix spectra, since they determine the value of $S$. In Fig.\ref{SPEC1}
spectra of the pure chain for fixed length and five different values of $\Delta$
are shown on the left. Plotted are the 100 largest eigenvalues $w_n$ of $\rho$,
ordered according to their magnitude. One sees the typical initial exponential decrease 
which then flattens out. A variation of $\Delta$ leads only to relatively small
changes in the curves and $S$ will also not be affected much. Note, however,
that $S$ is mainly determined by the first few large eigenvalues.
On the right,
the analogous spectrum is plotted for a defect with strength $J_{imp}=0.5$. Here one can
see a marked variation with $\Delta$. The decay is fastest for $\Delta=-0.9$
and slowest for $\Delta=+0.9$. Because of the sum rule $\sum w_n = 1$,
a slower decay is coupled with a smaller value for the first or first few eigenvalues
\cite{PZhao05}. This can be seen in the inset of the figure.\\
The size dependence of the spectra is shown in Fig.\ref{SPEC2} for two values of
$\Delta$. In both cases the curves become flatter as $L$ increases, which is 
well-known from other DMRG calculations. The behaviour of the first eigenvalues, 
however, is different. For $\Delta=-0.9$ one has a region of relatively slow
initial decay which becomes even slower as $L$ increases. The first two 
eigenvalues decrease in accord with the remark made above. For $S$ one therefore
expects a relatively large value, because several $w_n$ are of order 1,
and a marked size dependence.
For $\Delta=+0.9$, on the other hand, there is a gap between the first and the second
eigenvalue and no visible change with $L$. Because of the rather fast
decay of the complete spectrum, the sum rule is not effective here. Moreover, the
largest eigenvalue $w_0$ is close to 1 and thus gives only a small contribution to $S$.
Thus one expects a small absolute value and a weak size dependence of $S$ here.

\section{Entanglement entropy}

\begin{figure}
\includegraphics[width=6.8cm]{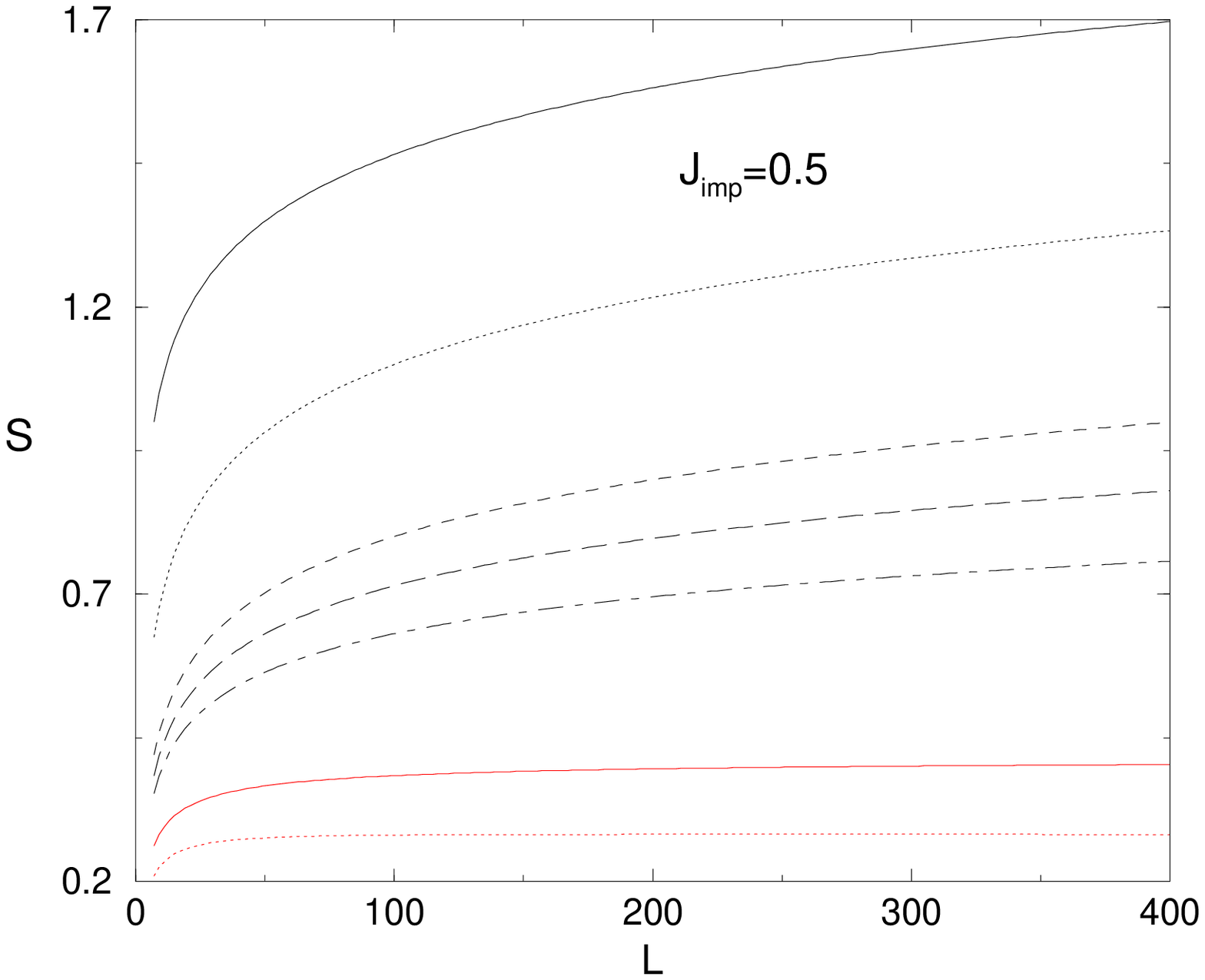}
\includegraphics[width=6.8cm]{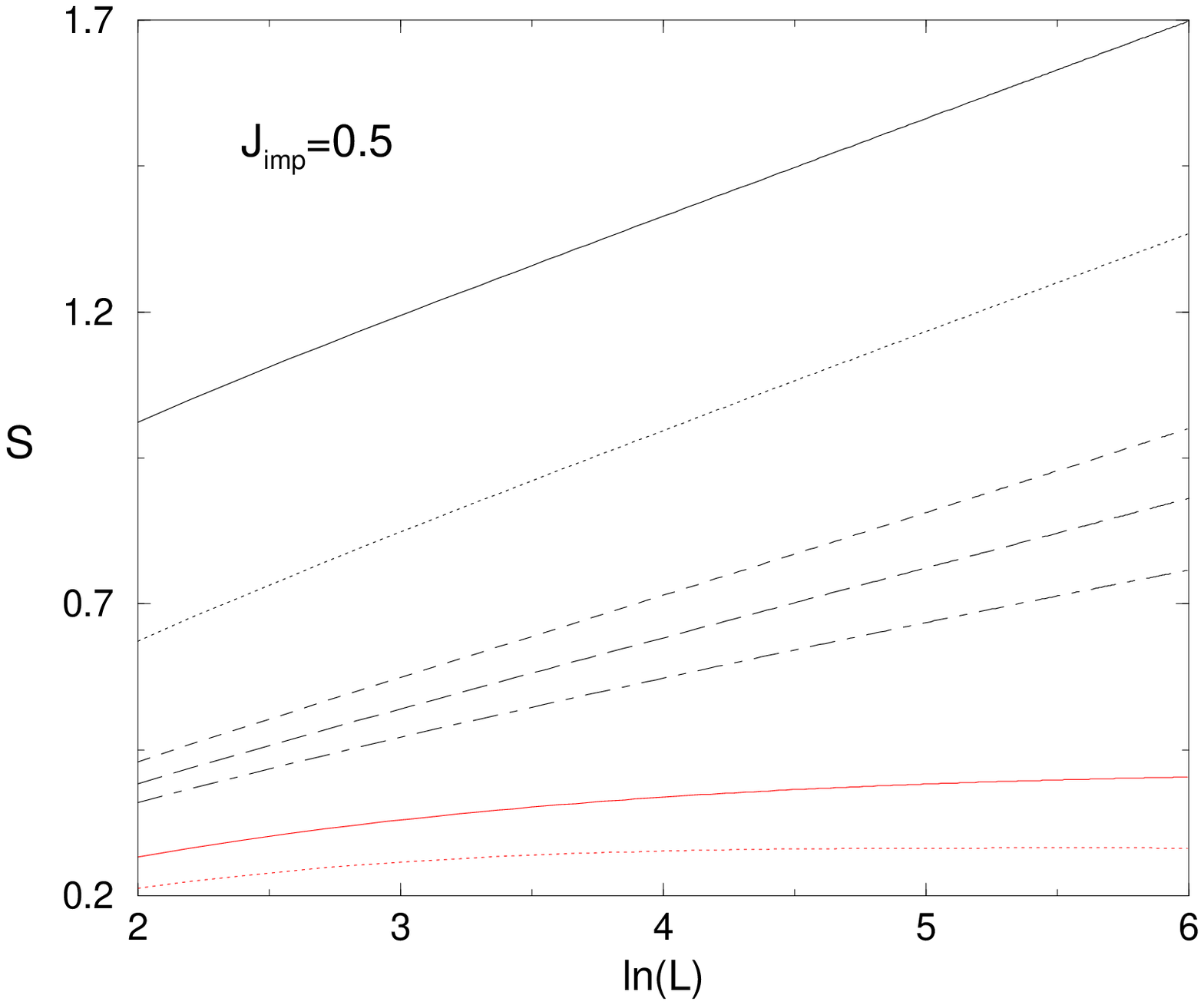}
\caption{Entanglement entropy as function of the system size for $J_{imp}=0.5$
and seven values of the anisotropy. The data correspond to $\Delta= -0.9;-0.5;
-0.1;0;0.1;0.5;0.9$, from top to bottom.}
\label{ENTRO}
\end{figure}

\begin{figure}
\includegraphics[width=6.8cm]{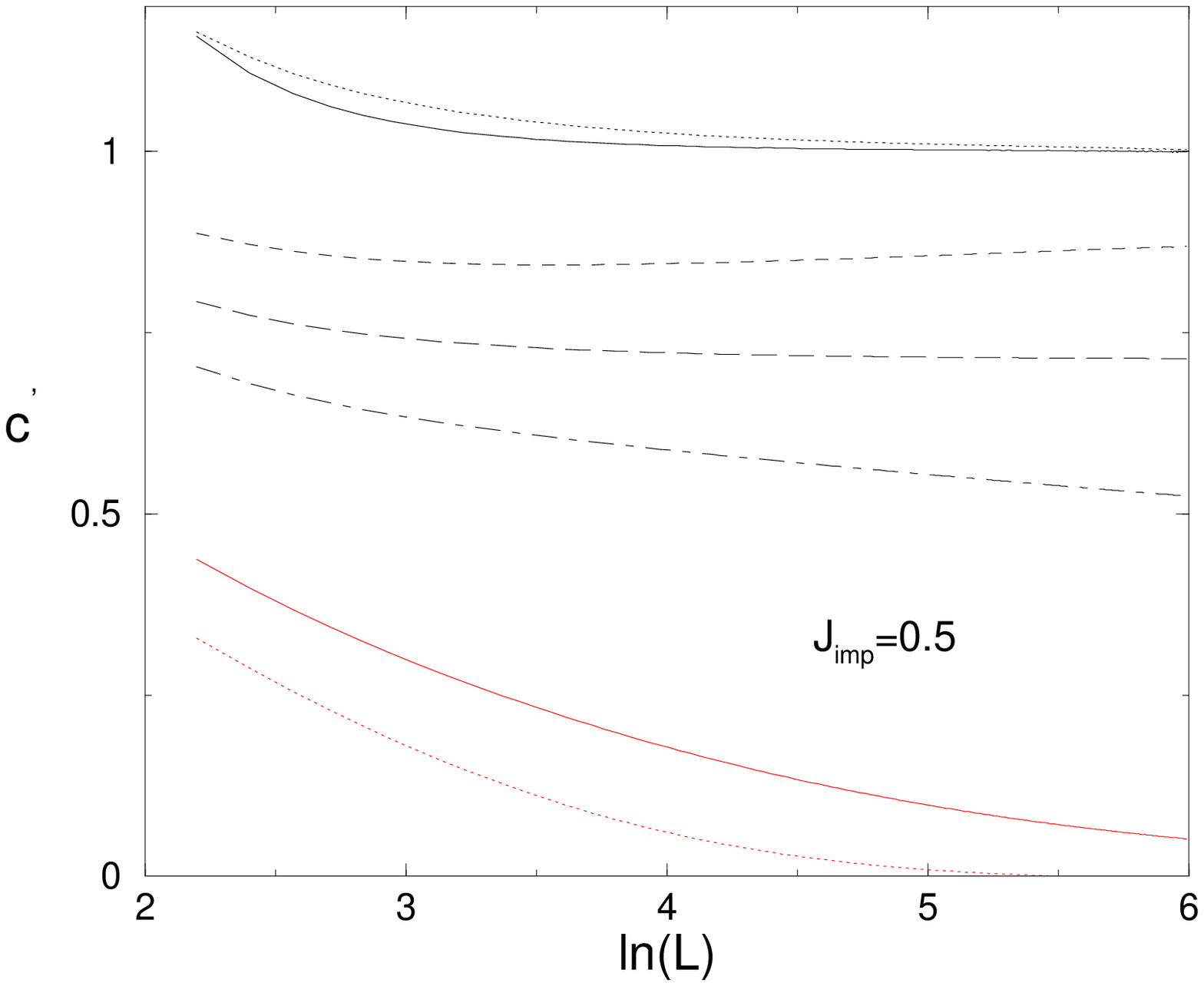}
\includegraphics[width=6.8cm]{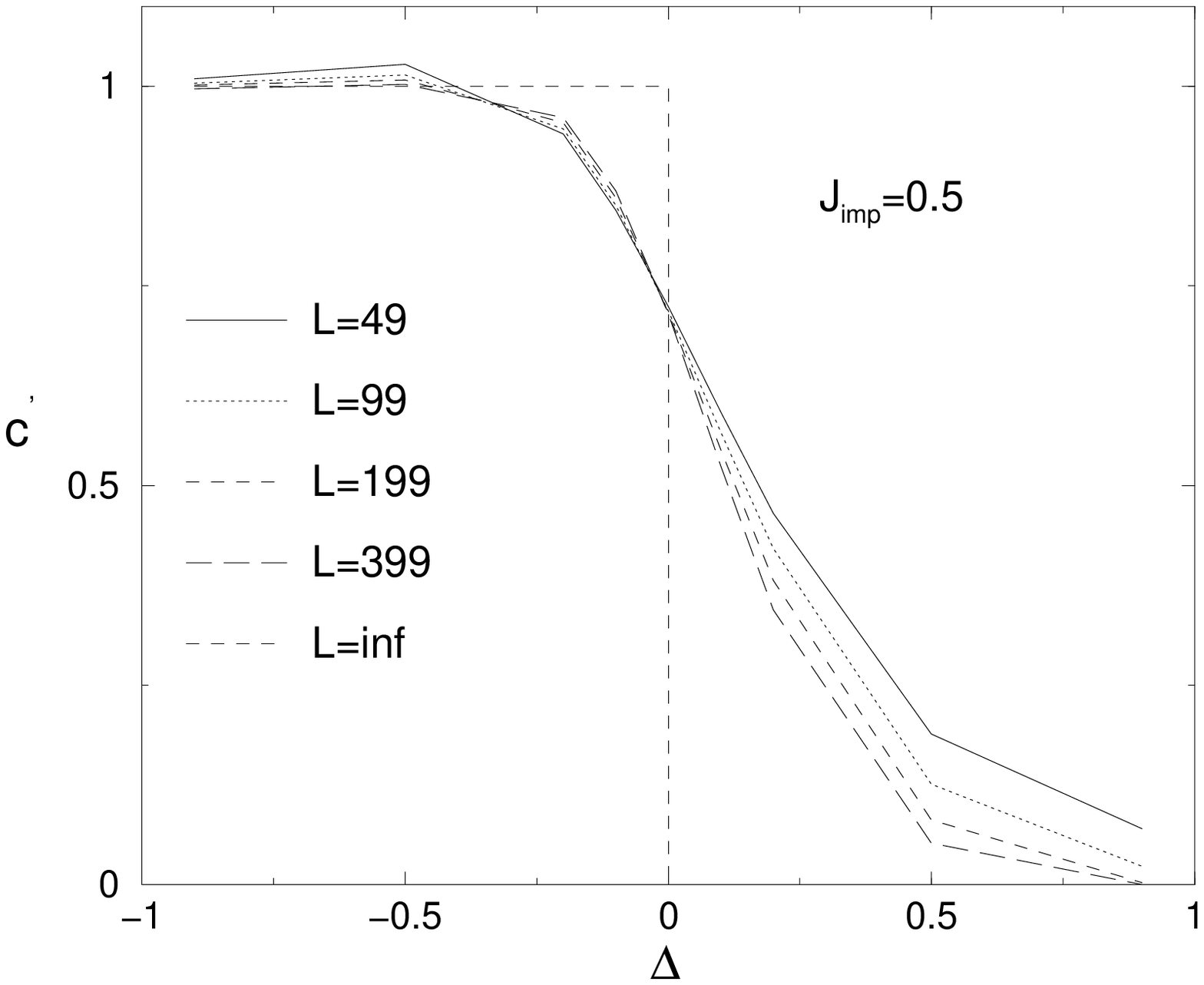} 
\caption{Effective central charge for $J_{imp}=0.5$. Left: as function of the system size 
for $\Delta= -0.5;-0.9;-0.1;0;0.1;0.5;0.9$, from top to bottom.
Right: as function of the anisotropy.}
\label{CEFF05}
\end{figure}

Results for the entropy $S(L)$ are shown in Fig.\ref{ENTRO} for $J_{imp}= 0.5$
and sizes up to $L=401$. One can see a monotonic increase with $L$ for all values
of $\Delta$. The absolute value of $S$ and the dependence on $L$ are large for
negative $\Delta$ and small for positive $\Delta$, confirming the expectations
based upon the spectra. The plot against $\ln L$ on the right hand side
shows relatively straight curves with similar slopes in the first case but rather
bent curves which seem to become horizontal in the second one. Thus there are
characteristic differences between repulsive and attractive interactions. 
A determination of the slope and the effective central charge via (\ref{ceff})
leads to the results shown in Fig. \ref{CEFF05}. The quantity $c'$ clearly tends to 
the value 1 of the pure system for large negative $\Delta$ and to zero for large
positive ones. The data for $\Delta= \pm 0.1$ also seem to fit into the pattern,
while for $\Delta=0$ the limit obviously differs from 1. This corresponds to the
results in \cite{Peschel05} where a continuous variation of the effective central
charge with the defect strength was found in this case. For a detailed comparison, one 
has to take into account that the quantity $c_{eff}$ used there referred to a subsystem
coupled to the rest with e.g one modified and one unmodified bond. The 
contributions of these two bonds add up and the relation to $c'$ is therefore
\be
c_{eff}/3 = c'/6  + 1/6
\label{ceff1}
\ee
In \cite{Peschel05} the asymptotic value of $c_{eff}$ was determined  by an extrapolation
in $1/L$. The same procedure also works here and the results agree to 3-4 decimal places.
In view of the rather different geometries in the two cases, this is a non-trivial
consistency check of the calculations.\\

\begin{figure}
\includegraphics[width=6.8cm]{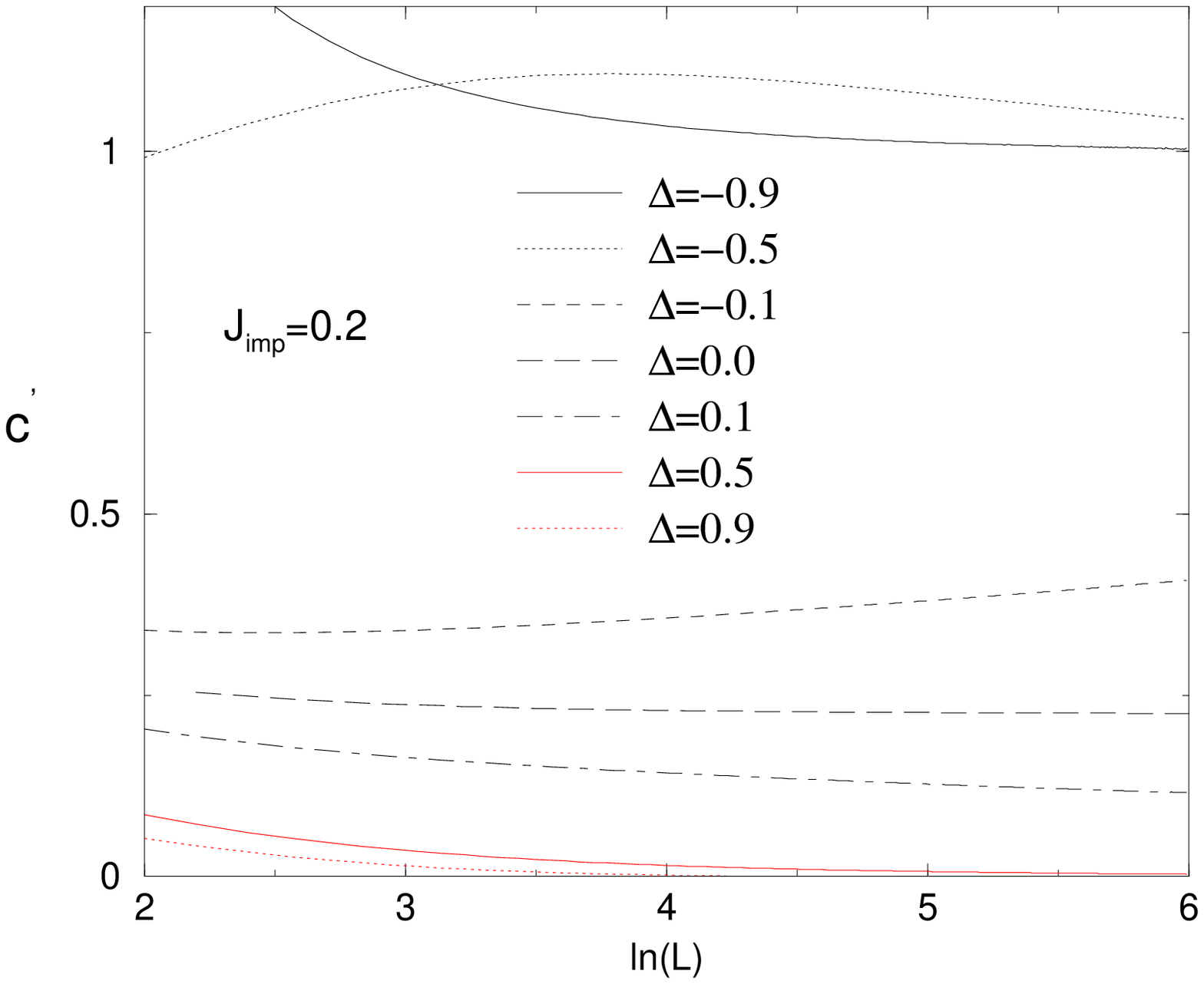}
\includegraphics[width=6.8cm]{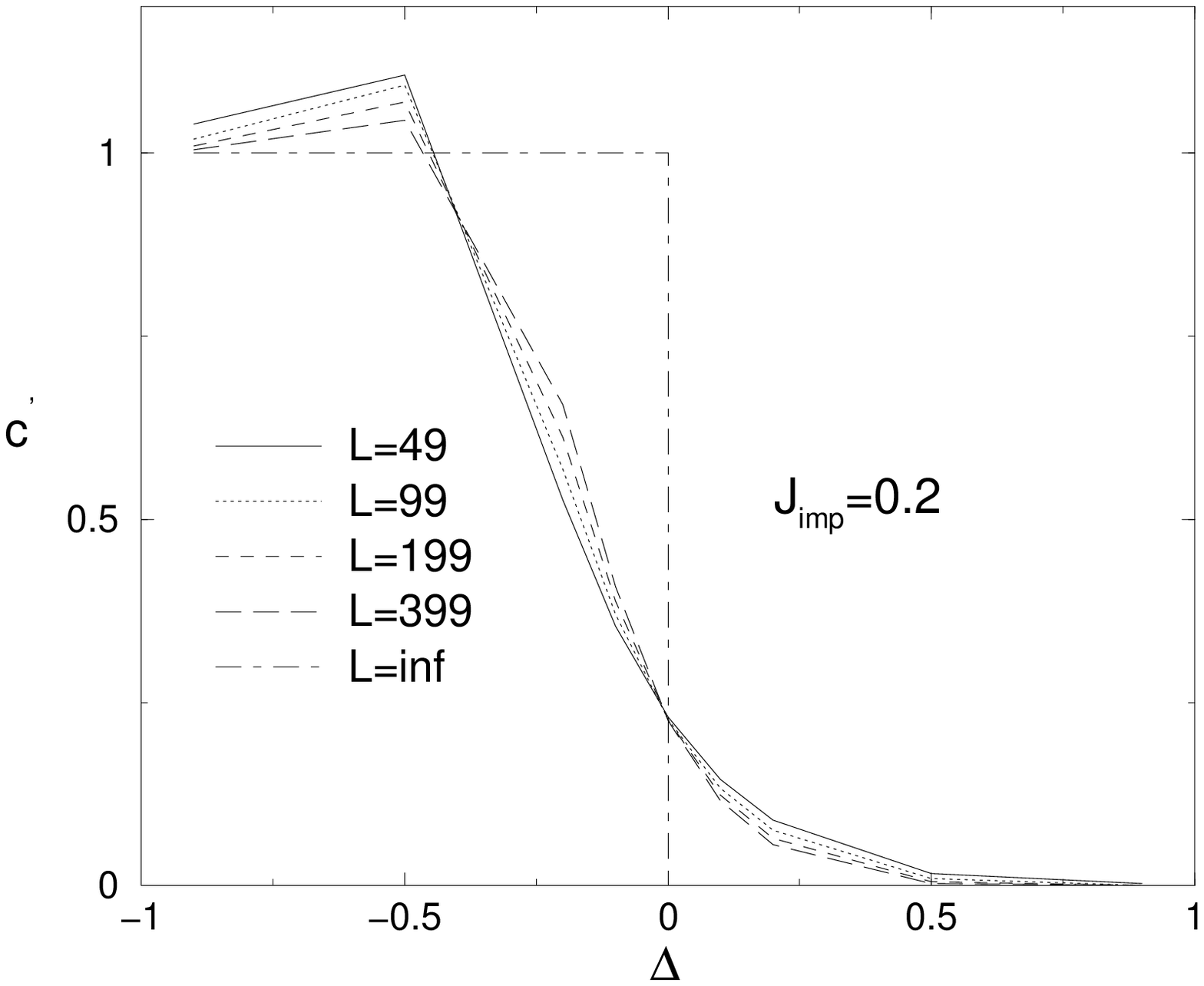} 
\caption{The same as Fig. \ref{CEFF05}, but for $J_{imp}=0.2$. }
\label{CEFF02}
\end{figure}

The right part of Fig.\ref{CEFF05} gives $c'$ as a function of $\Delta$ and shows
the different behaviour for positive and negative values very clearly. To show how
these effects vary with the impurity strength, we present results for two other
values of $J_{imp}$. In Fig. \ref{CEFF02} this is done for a weaker bond, 
corresponding to a stronger perturbation. Then $c'$ for the non-interacting
case is only about 0.25 and even smaller for repulsion. For weak attraction, the values
rise only slowly but for strong attraction they exceed 1 and approach it from above.

\begin{figure}
\includegraphics[width=6.8cm]{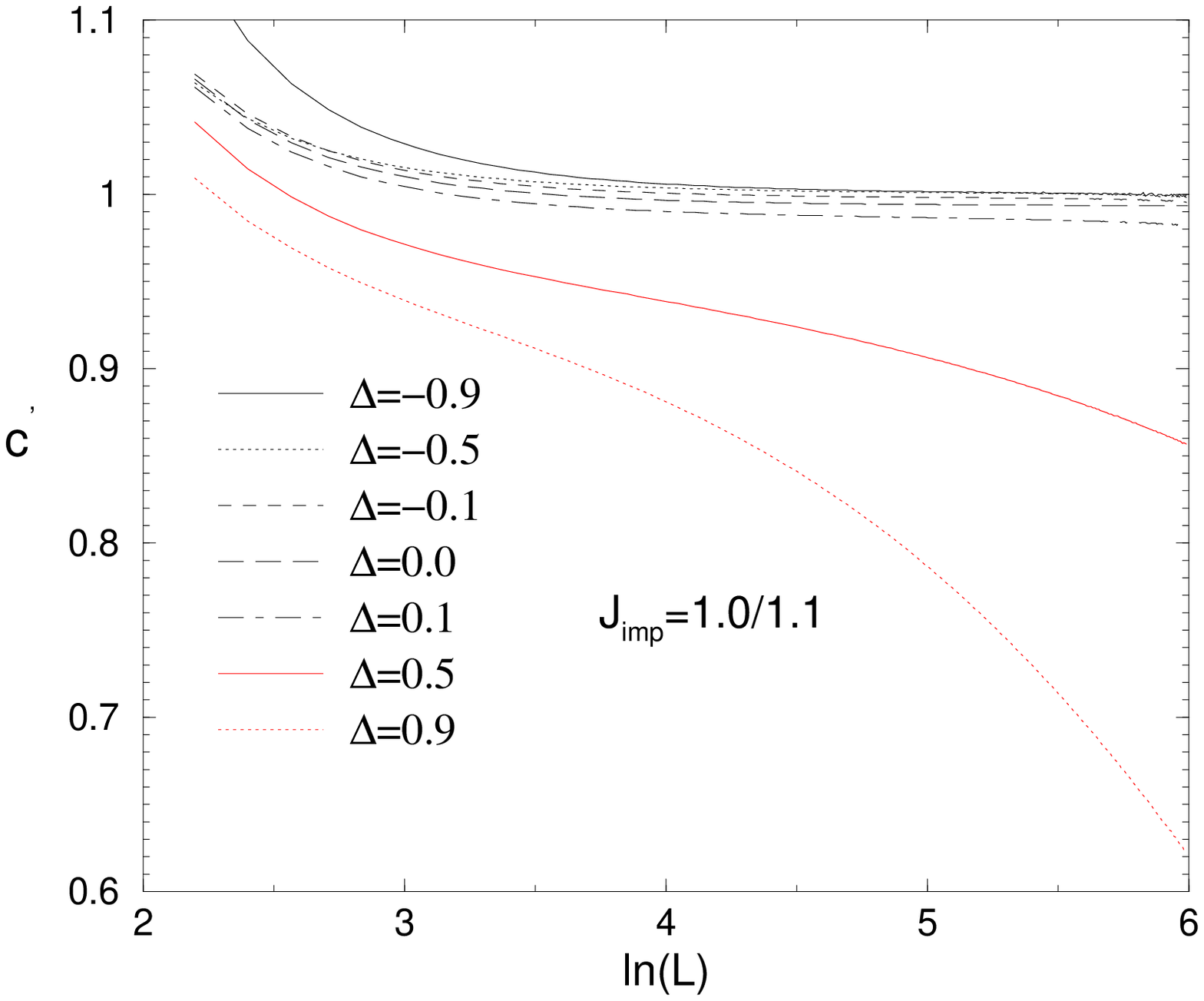}
\includegraphics[width=6.8cm]{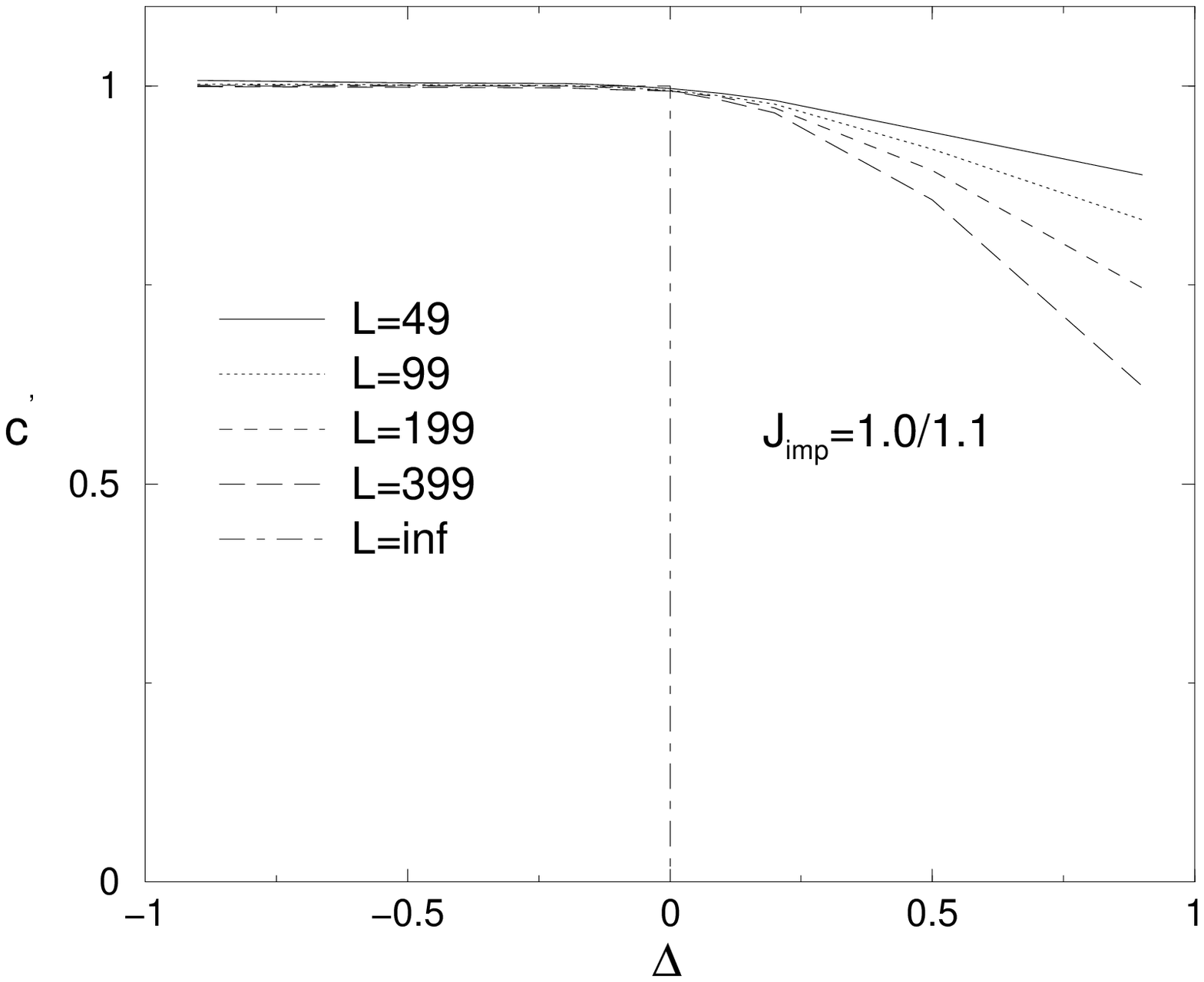}
\caption{The same as Fig. \ref{CEFF05}, but for $J_{imp}=1/1.1=0.9091$. Please note the
different vertical scale in the left figure.}
\label{CEFF09}
\end{figure}

In Fig.\ref{CEFF09}, on the other hand, the results for an only slightly weakened
bond, $J_{imp}=1/1.1=0.9090..$, are shown. In this case, not only the $c'$ values for 
negative $\Delta$ lie rather close to 1, but also those for $\Delta=0, 0.1$. Only
for the larger $\Delta$ one has a reduction below 1, but the values are still rather
high. From the left figure one sees that the curves for these $\Delta$  bend downwards
and presumably approach zero for longer chains. However, the asymptotic region is beyond
the range of our system sizes.  

\section{Discussion}

These results show a clear distinction between attractive and repulsive
interactions and a crossover between the two which is determined by the strength of
the perturbation. This strength fixes the value of $c'$ in the non-interacting
system. Because this value depends only weakly on the size, it serves as a reference
point through which the $c' vs.\Delta$ curves for different sizes go. The curves, which we
determined only at a number of points, will be continuous in all
finite systems.
These findings are in line with the general predictions on the influence of an
impurity in a Luttinger liquid, namely that for attraction the perturbation 
heals with increasing system size, while for repulsion it becomes stronger. For
the weakened bonds which we have considered, the latter case means a scaling
towards zero and a cutting of the chain. The vanishing of $c'$ is in agreement
with this picture, but the cutting should not be taken too literally since it only 
refers to processes near the Fermi energy. 
As Fig.\ref{ENTRO} shows, there remains a residual entanglement
corresponding to the constant $k$ in (\ref{log}) which depends on the interaction 
strength. \\

\begin{figure}
\includegraphics[width=6.8cm]{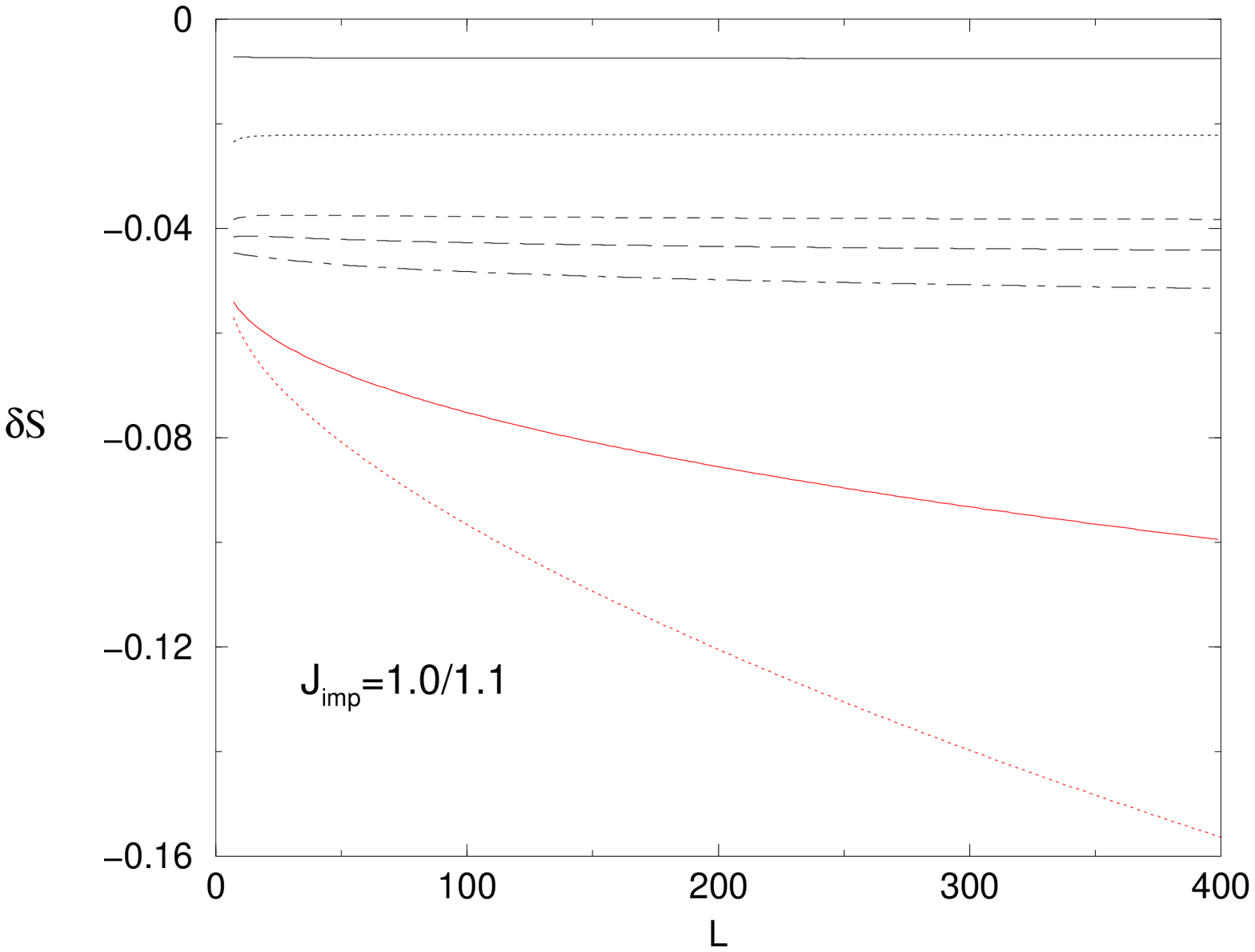}
\includegraphics[width=6.8cm]{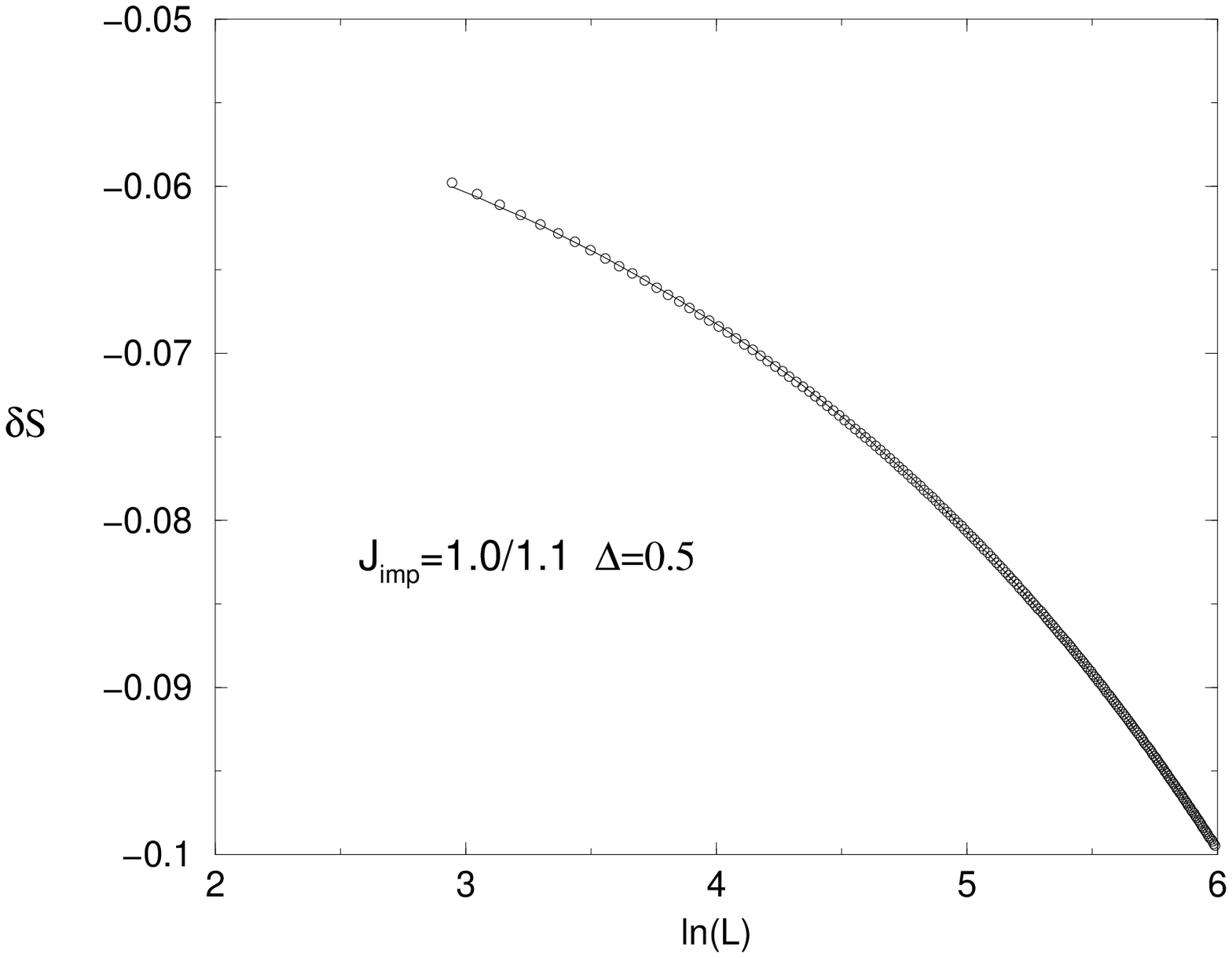}
\caption{Defect contribution $\delta S$ to the entanglement entropy for $J_{imp}= 1/1.1$. 
Left: $\delta S$ as function of $L$ for $\Delta= -0.9;-0.5;-0.1;0;0.1;0.5;0.9$, from top to
bottom. Right: Fit of the data (circles) for $\Delta= 0.5$ according to eqn.(\ref{levine}), see
text.}
\label{DELTA-S}
\end{figure}

There is no general expression for $S$ in a finite system, not even for the XX case, 
but in \cite{Levine04}
a formula was given for a weak impurity in a Luttiger liquid. The contribution to
$S$ is then
\be
\delta S= -y^2\epsilon^2 \;\frac {g}{1-g^2}\;(\frac {L}{\epsilon})^{1-g}
\ln(\frac {L}{\epsilon}) - b(g)
\label{levine}
\ee
where $y$ is the strength of the impurity potential, $\epsilon$ a small-distance cutoff,
$b(g)$ a positive constant and $g$ the Luttinger liquid parameter measuring the
interaction. For the XXZ lattice model, it can be expressed as
\be
\frac {1}{g} = 2\; (1-\frac {1}{\pi} \rm arccos \,\Delta )
\label{luttinger}
\ee
For attraction ($\Delta < 0$) one has $g > 1$ and for repulsion $g < 1$. In the latter
case, $\delta S$ is always negative and diverges with $L$. The numerical results for
$\delta S$ on the left of Fig.\ref{DELTA-S} actually show this behaviour. One can also
fit the data to equ.(\ref{levine}) as is shown on the right for 
$\Delta= 0.5$. The value of $g$ is then 0.75 and the fit gives $\epsilon = 13$ and 
$b=0.06$. However, similar fits for $\Delta = 0.1$ and $0.9$ lead to very different results
for $\epsilon$ and if one interprets it as an effective length determined by the
scaling equations there are still inconsistencies. For attraction the $L$-dependent term
in (\ref{levine}) is positive and decreases with $L$ but the data for $\delta S$ are
more or less $L$-independent and cannot be fitted with reasonable parameters. Thus
we cannot verify Levine's formula in detail. One should mention that he also predicts
a $\ln^2 L$ term in $\delta S$ for the non-interacting system which one does not see
in the numerics. It could be that he used a too simple geometry for the two-dimensional
bosonic field theory in his approach.\\
Summing up, we have shown by numerical calculations that the general effects of an 
impurity in a Luttinger liquid also show up in the entanglement entropy. Its logarithmic
behaviour remains for attraction but is suppressed for repulsion. A more detailed 
analytical treatment is, however, still desirable.\\

{\it Acknowledgement :} I.P. thanks the Interdisciplinary Center for Theoretical Studies, 
CAS, Beijing, for its hospitality. He also acknowlegdes discussions with N. Shannon and 
V. Meden and correspondence with N. Laflorencie.

\end{document}